\documentclass{article}
\usepackage{spconf,amsmath,graphicx}
\usepackage{xcolor}

\usepackage{enumitem}
\setlist{nosep, leftmargin=14pt}

\usepackage{mwe} 
\usepackage{amsfonts}

\title{Analyzing Model Misspecification in Quantitative MRI: Application to Perfusion ASL
\thanks{\scriptsize{\copyright\ 2026 IEEE. Personal use of this material is permitted. Permission from IEEE must be obtained for all other uses, in any current or future media, including reprinting/republishing this material for advertising or promotional purposes, creating new collective works, for resale or redistribution to servers or lists, or reuse of any copyrighted component of this work in other works.}}
}
\name{Jiachen Wang$^{\dagger}$, Jonathan I.\ Tamir$^{*}$, Adam Bush$^{\dagger}$}
\address{$^{\dagger}$ The University of Texas, Austin, BME $^{*}$ The University of Texas, Austin, ECE}
\begin{document}
%
\maketitle
\begin{abstract}
Quantitative MRI (qMRI) involves parameter estimation governed by an explicit signal model. However, these models are often confounded and difficult to validate \emph{in vivo}. A model is misspecified when the assumed signal model differs from the true data-generating process. Under misspecification, the variance of any unbiased estimator is lower-bounded by the misspecified Cramér–Rao bound (MCRB), and maximum-likelihood estimates (MLE) may exhibit bias and inconsistency. Based on these principles, we assess misspecification in qMRI using two tests: (i) examining whether empirical MCRB asymptotically approaches the CRB as repeated measurements increase; (ii) comparing MLE estimates from two equal-sized subsets and evaluating whether their empirical variance aligns with theoretical CRB predictions. We demonstrate the framework using arterial spin labeling (ASL) as an illustrative example. Our result shows the commonly used ASL signal model appears to be specified in the brain and moderately misspecified in the kidney. The proposed framework offers a general, theoretically grounded approach for assessing model validity in quantitative MRI.

\end{abstract}

\section{Introduction}
\label{sec:intro}
Quantitative MRI (qMRI) relies on a biophysical model that links the acquired signal to physiological parameters. Therefore, accurate parameter estimation depends on the validity of the signal models. Ideally, models are validated using an independent clinical standard test, which may be unavailable in some applications. In statistical learning, a model is said to be \textit{misspecified} when the assumed signal-generation process differs from the true data-generating distribution \cite{fortunati2017performance}. Model misspecification can induce systematic bias and increased variance in parameter estimates, which in turn undermines the reliability and interpretability of qMRI.

Model validity in qMRI has been explored through two general directions. First, goodness-of-fit–based metrics evaluate how well the model explains the data using residual analysis, split-data reproducibility, or sensitivity of estimates to fixed parameters \cite{loh2008residual}. These methods can reveal poor fits but do not provide a statistical measure of how model mismatch affects estimator variance or consistency. Second, model-refinement approaches introduce extended or multi-compartment biophysical models to improve physiological plausibility \cite{parkes2002improved}. While these approaches reduce mismatch, they do not evaluate how misspecified the original model is, and whether the proposed model is truly better specified.

A complementary statistical framework for assessing misspecification is the  misspecified Cramér–Rao bound (MCRB) \cite{white1982maximum,vuong1986cramer}, which generalizes the conventional Cramér–Rao bound (CRB) to account for potential model mismatch.
While the MCRB has been used in economics \cite{cerreia2025making} and wireless communications \cite{abed2021misspecified}, it has not been applied as a tool in qMRI. Our work fills this gap by introducing a practical MCRB-based framework that provides a quantitative misspecification metric and corresponding voxelwise maps.


When a model is misspecified, two consequences follow: (i) the minimum achievable variance of an unbiased estimator is limited by the MCRB, which provides a strictly looser bound than the conventional CRB \cite{vuong1986cramer}; (ii) under maximum-likelihood estimation (MLE), parameter estimates can exhibit data-dependent biases and inconsistency \cite{white1982maximum}. Misspecification may be due either to incorrect fixed parameters of the model or to an incorrect parametric model itself \cite{fortunati2017performance}.

Guided by the theory above, we propose two tests for assessing misspecification. First, we evaluate the \emph{asymptotic convergence} of empirical MCRB and whether it approaches empirical CRB as the number of repeated measurements increases. Second, we evaluate \emph{parameter estimate consistency} through a comparison of the estimates from two equal-sized subsets of the measured data, and assessing the gap between their empirical variances and the theoretical CRB.
We also assess the degree of misspecification attributable to incorrect fixed parameters.

While our proposed framework described applies generally to qMRI, we demonstrate its use in arterial spin labeling (ASL) as an example. ASL is a noninvasive, non-contrast technique that magnetically labels blood water to quantify perfusion: microvascular delivery of blood to exchanging capillaries. The Buxton general kinetic model is the most widely used method for describing the ASL signal and converting measured perfusion-weighted images into quantitative perfusion (mL/min/100g) \cite{alsop2015recommended}.

\section{Theory}
\label{sec:theory}
\subsection{qMRI Forward Model}
    Without loss of generality, under prewhitened Gaussian noise and Nyquist sampling, each qMRI voxel is distributed as
    \begin{equation}
        \begin{aligned}
        \mathbf{x}^{(m)} 
        \sim f(\mathbf{x}; \boldsymbol{\theta}, \mathbf{u}) 
           &= \mathcal{N}\!\left(
                \mathcal{F}(\boldsymbol{\theta}, \mathbf{u}),
                \sigma^2 \mathbf{I}_{N\times N}
             \right), \\
        &\quad\quad\quad\quad\quad\quad m = 1,\dots,M,
        \end{aligned}
        \label{eq:gauss}
    \end{equation}
    where $\mathbf{x}^{(m)}\in\mathbb{C}^N$ are the measurements from the $m^{\mathrm{th}}$ experiment (called dynamics, averages, repetitions, NEX, etc.), $\boldsymbol{\theta} \in \mathbb{R}^P$ are the fixed but unknown physiological parameters, $\mathbf {u}\in\mathbb{R}^N$ are user controllable scanner parameters, $\mathcal{F}$ denotes the discretized signal model evaluated at $\mathbf u$, and $\sigma^2$ is the noise variance. When magnitude-only reconstructions are used, the measurements are Rician distributed, but the noise is well approximated as Gaussian when the signal-to-noise ratio (SNR) is sufficiently high \cite{gudbjartsson1995rician}. 
    
    \subsubsection{ASL Buxton Model}
        The Buxton model treats each imaging voxel as a single, well-mixed compartment in which labeled water arrives after an arterial transit time (ATT) with a perfusion rate $f$, and undergoes longitudinal relaxation characterized by the tissue $T_{1}$. It models the perfusion-weighted signal $\Delta M(t)$ in the following convolutional form \cite{buxton1998general}:
        \begin{align}
            \Delta M(t)
            &= 2\alpha f M_{0b}\int_{0}^{t} c(\tau)\,r(t-\tau)\,m(t-\tau)
              \,\mathrm{d}\tau \notag \\
            &=2\alpha f M_{0b}\left\{c(t)*\left[r(t)\,m(t)\right]\right\},
            \label{eq:buxton}
        \end{align}
        where the factor $\alpha$ accounts for the labeling efficiency and $M_{0b}$ denotes the equilibrium magnetization of arterial blood. The residue and magnetization relaxation functions are
        \begin{align}
            r(t)&=\exp\bigl(-ft/\lambda\bigr), \quad m(t)=\exp\bigl(-t/T_{1}\bigr),
        \label{eq: rt_mt}
        \end{align}
        where $\lambda$ is the blood-tissue partition coefficient. The arterial input function $c(t)$ is modeled as a bolus of duration $\tau$ that arrives at time ATT:
        \begin{equation}
            c(t)=
            \begin{cases}
            \exp \left(- (t-\mathrm{ATT}) / T_{1b} \right), 
            & \mathrm{ATT} < t < \mathrm{ATT} + \tau,\\[0.4em]
            0, & \text{otherwise},
            \end{cases}
            \label{eq:c_t}
        \end{equation}
        where $T_{1b}$ denotes the blood longitudinal relaxation time.
        
        To fit the general framework of Eq.~[\ref{eq:gauss}], $\boldsymbol{\theta} = [f, \mathrm{ATT}]^{\top}$, $\mathbf{u}$ is the vector of sampled time points (called post-labeling delay, PLD), and $\mathcal{F}$ is the Buxton model of Eqs.~[\ref{eq:buxton}-\ref{eq:c_t}].


\subsection{Misspecified Cramér–Rao Bounds}

    Let $p(\mathbf{x})$ denote the true data-generating distribution.
    When the assumed likelihood $f(\mathbf{x};\boldsymbol{\theta},\mathbf{u})$
    is correctly specified, the covariance of any unbiased estimator
    $\hat{\boldsymbol{\theta}}$ is bounded below by the CRB:
    \begin{equation}
            \mathbf{C}_{\mathrm{C}} = -\mathcal{A}(\boldsymbol{\theta})^{-1},
            \label{eq:crb}
    \end{equation}
    where $\mathcal{A}(\boldsymbol{\theta})$ is the 
    \emph{expected Hessian} of the log-likelihood,
    \begin{equation}
        \mathcal{A}(\boldsymbol{\theta}) = \mathbb{E}_p\left[\frac{\partial^2}{\partial \boldsymbol{\theta}^2}\log f(\mathbf{x}; \boldsymbol{\theta}, \mathbf{u})\right].
        \label{eq:a}
        \end{equation}
    However, if $f(\mathbf{x};\boldsymbol{\theta},\mathbf{u})$ does not coincide with $p(\mathbf{x})$, the CRB becomes overly optimistic. In this general case, the relevant lower bound is the MCRB, given by \cite{white1982maximum,vuong1986cramer}
    \begin{equation}
        \mathbf{C}_{\mathrm{M}}
        = \mathcal{A}^{-1}(\boldsymbol{\theta})\, \mathcal{B}(\boldsymbol{\theta})\, \mathcal{A}^{-1}(\boldsymbol{\theta}),
        \label{eq:mcrb}
    \end{equation}
    where
    \begin{equation}
        \mathcal{B}(\boldsymbol{\theta}) = \mathbb{E}_p\!\left[\frac{\partial \log f(\mathbf{x}; \boldsymbol{\theta}, \mathbf{u})}{\partial \boldsymbol{\theta}}
            \frac{\partial \log f(\mathbf{x}; \boldsymbol{\theta}, \mathbf{u})^{\top}}{\partial \boldsymbol{\theta}}\right].
    \end{equation}
    
    When the model is correctly specified,  $\mathbf{C}_{\mathrm{M}} = \mathbf{C}_{\mathrm{C}}$.
    In practice, $p(\mathbf{x})$ is unknown, so we approximate the expectations with empirical estimates $\hat{\mathbf{C}}_{\mathrm{M}}$ and $\hat{\mathbf{C}}_{\mathrm{C}}$, which converge to $\mathbf{C}_{\mathrm{M}}$ and $\mathbf{C}_{\mathrm{C}}$ as $M \rightarrow \infty$ (defined in \cite{fortunati2017performance} Eqs.~13-15).
    

\subsection{Evaluation Metrics}
    We compare $\hat{\mathbf{C}}_{\mathrm{M}}$ and $\hat{\mathbf{C}}_{\mathrm{C}}$ via the congruence transformation
    \begin{equation}
        \hat{\mathbf{P}}= \hat{\mathbf{C}}_{\mathrm{C}}^{-1/2} \hat{\mathbf{C}}_{\mathrm{M}} \hat{\mathbf{C}}_{\mathrm{C}}^{-1/2} = \mathbf{W}^{\top} \hat{\mathbf{C}}_{\mathrm{M}} \mathbf{W}, \quad \mathbf{W} \equiv \hat{\mathbf{C}}_{\mathrm{C}}^{-1/2},
    \end{equation}
    where the choice of $\mathbf{W}$ corresponds to whitening with respect to the CRB. If the model is correctly specified, $\hat{\mathbf{P}} \rightarrow \mathbf{I}_{P\times P}$.
    
    The eigenvalues of $\hat{\mathbf{P}}$ quantify the degree of variance amplification when comparing $\hat{\mathbf{C}}_{\mathrm{M}}$ to
    $\hat{\mathbf{C}}_{\mathrm{C}}$. In particular, the condition number $\kappa\;\equiv\; \lambda_{\max}/\lambda_{\min}$ summarizes how anisotropic this amplification is across parameter combinations. Under correct model specification, CRB and MCRB coincide, so that $\lambda_{\max}, \lambda_{\min}, \kappa \rightarrow 1$ as $M\rightarrow\infty$.

\section{Methods}

\begin{figure*}[t!]
   \centering
   \centerline{\includegraphics[width=\linewidth]{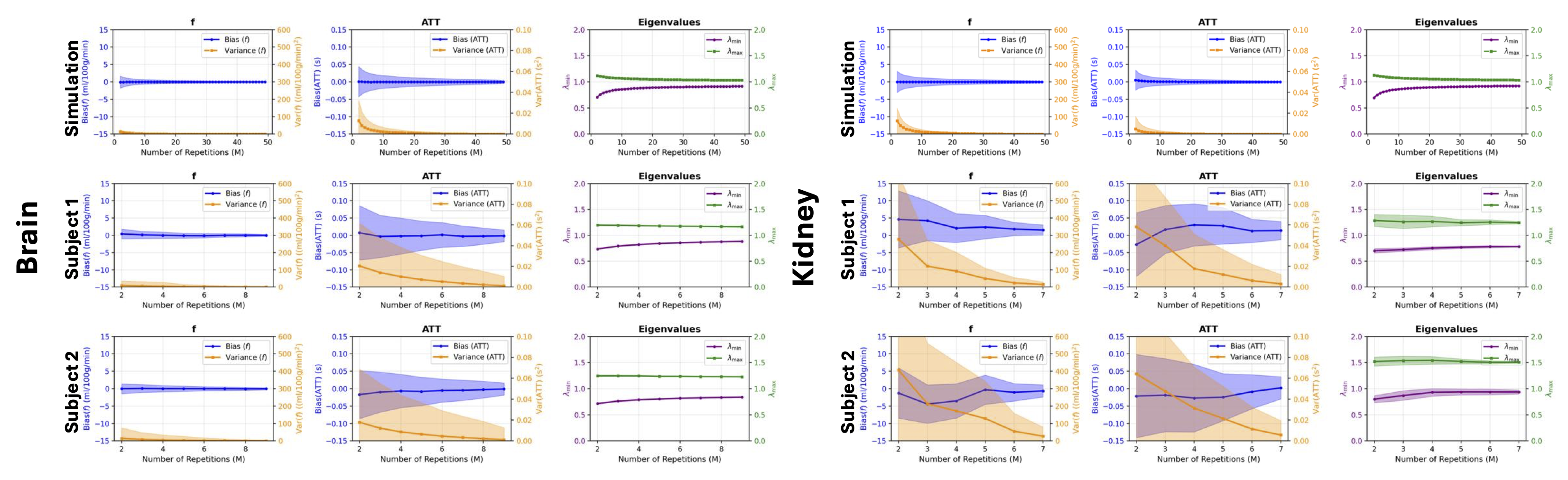}}
   \caption{Simulated and \emph{in vivo} asymptotic convergence as the number of repeated measurements increases. Subject 1 represents the better-conditioned case for both organs. In simulation, as $m$ becomes sufficiently large, all metrics converge to their theoretical values. \emph{In vivo}, the eigenvalues $\lambda_{\max}$, $\lambda_{\min}$ do not converge to 1 as $m$ increases, especially in the kidney.} 
   \label{fig:converge}
 \end{figure*}
\label{sec:method}
\subsection{\emph{in vivo} ASL Scans}
    Under Institutional Board Approval and informed consent, two healthy volunteers were scanned using pseudo-continuous ASL (pCASL) on a 3T Siemens Vida scanner for the brain and kidney, respectively. Acquisition parameters were: $N=21$ PLDs (0–1 s step 0.1 s, 1–3 s step 0.2 s), bolus duration $\tau = 1.5$s, repeated measurements $M_{\mathrm{total}} = 10$ for the brain, and $M_{\mathrm{total}} = 8$ for the kidney. During kidney imaging, subjects were instructed to maintain a paced breathing pattern to mitigate motion artifacts. A fixed global tissue $T_1 = 1.2$~s for brain and $T_1 = 1.4$~s for kidney was assumed \cite{alsop2015recommended}. Scanner-reconstructed magnitude images were used for all subsequent data processing.
    
\subsection{Asymptotic Convergence}
    To quantify the uncertainty induced by potential misspecification, we compute empirical CRB and MCRB from repeated measurements (see \cite{fortunati2017performance} Eqs.~13-15) and examine how they evolve as the number of repetitions increases. We first validate the procedure under a correctly specified setting by simulating the Buxton model for $M_{\mathrm{total}} = 50$ repetitions on 8{,}000 voxels uniformly spanning physiologically plausible parameter ranges ($f$ range: 0–150 mL/min/100g for the brain, 0–900 mL/min/100g for the kidney; ATT range 0–2 s \cite{alsop2015recommended,odudu2018arterial}) and adding Rician noise with variance matched to the \emph{in vivo} scans. For each $M \in \{2,\dots, M_{\mathrm{total}}\}$, we estimate the parameters via MLE, and then compute the bias, variance, $\lambda_{\max}$, and $\lambda_{\min}$ using a bootstrap with $K=10$ realizations. Each bootstrap resamples the $M$ repetitions (fixed PLDs), refits voxelwise MLE maps, and recomputes the metrics. The same procedure is applied to the \emph{in vivo} datasets. Because ground truth is unavailable, the parameter estimates obtained from all $M_{\mathrm{total}}$ repetitions are used as the reference when computing bias. The noise variance is estimated from background regions of the perfusion-weighted images.
        
\subsection{Subset Parameter Estimate Consistency}
    The ASL datasets are partitioned into two equal-sized subsets with approximately exponential spacing that emphasized either early (Set 1) or late (Set 2) PLDs. For each subset, we fit parameter maps with MLE, report the magnitude of relative error between the two estimates, and compute empirical parameter variances over $K=10$ bootstrap realizations for each $M$. These empirical variances are compared with the theoretical CRB in Eq.~[\ref{eq:crb}]. 
    
\subsection{Fixed Parameter Misspecification}
    To assess the degree of model misspecification due to fixed-parameter errors, we replace the global tissue $T_1$ with a tissue-specific $T_1$ map. Specifically, we assume two perfusion-related tissue types within each organ and assign a distinct $T_{1}$ value to voxels in the top 10\% of perfusion-weighted signal intensity. All acquisition parameters, PLDs, and estimation procedures are kept identical. We then compare the resulting $\lambda_{\max}$, $\lambda_{\min}$, and $\kappa$ against the global $T_1$ baseline.

  \begin{figure}[htbp]
   \centering
   \centerline{\includegraphics[width=\linewidth]{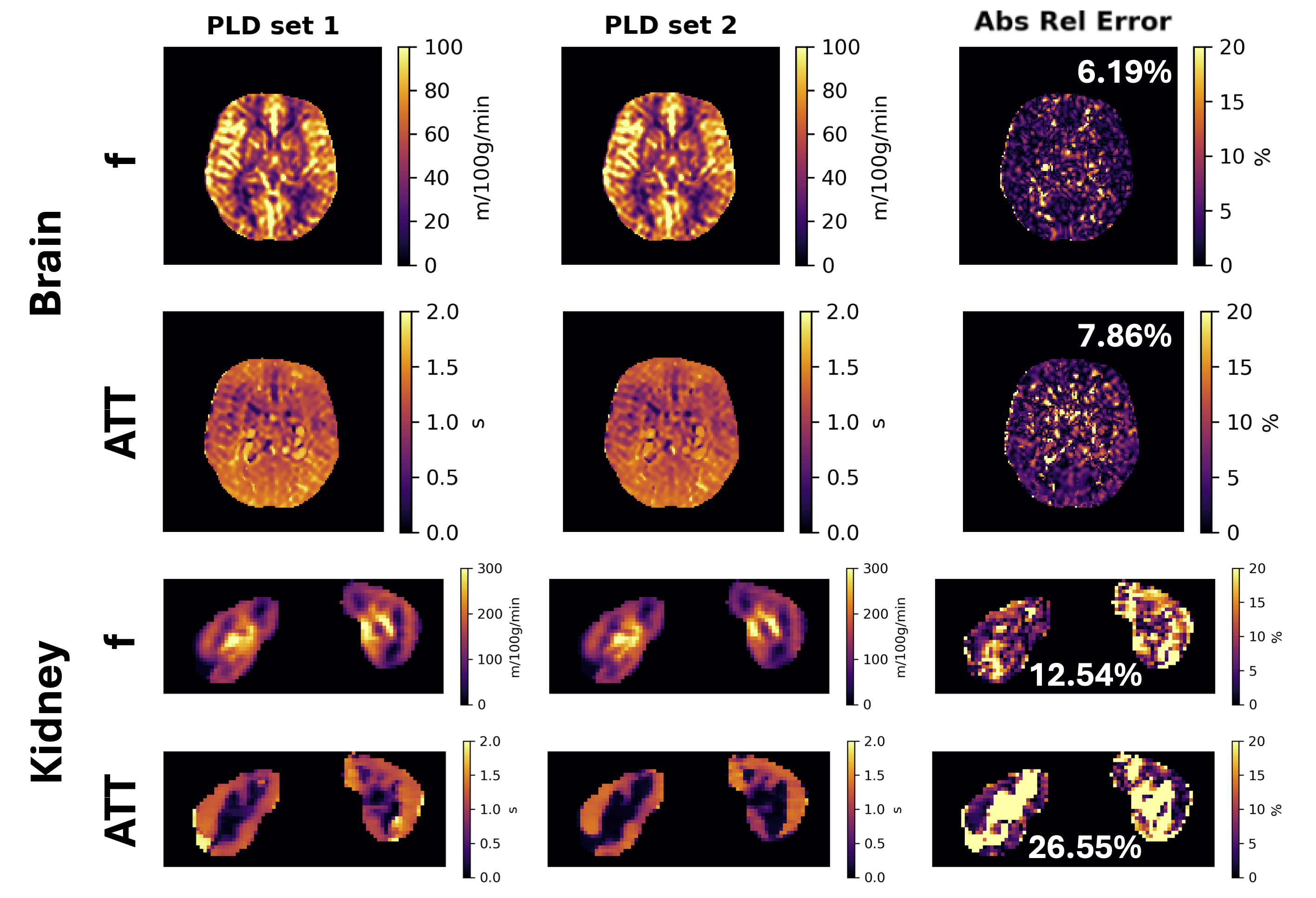}}
   \caption{Parameter estimates from two PLD subsets and their average relative error over all voxels (white text). Errors in the kidney are larger and more spatially heterogeneous.} 
   \label{fig:consist}
 \end{figure}

  \begin{figure}[htbp]
   \centering
   \centerline{\includegraphics[width=\linewidth]{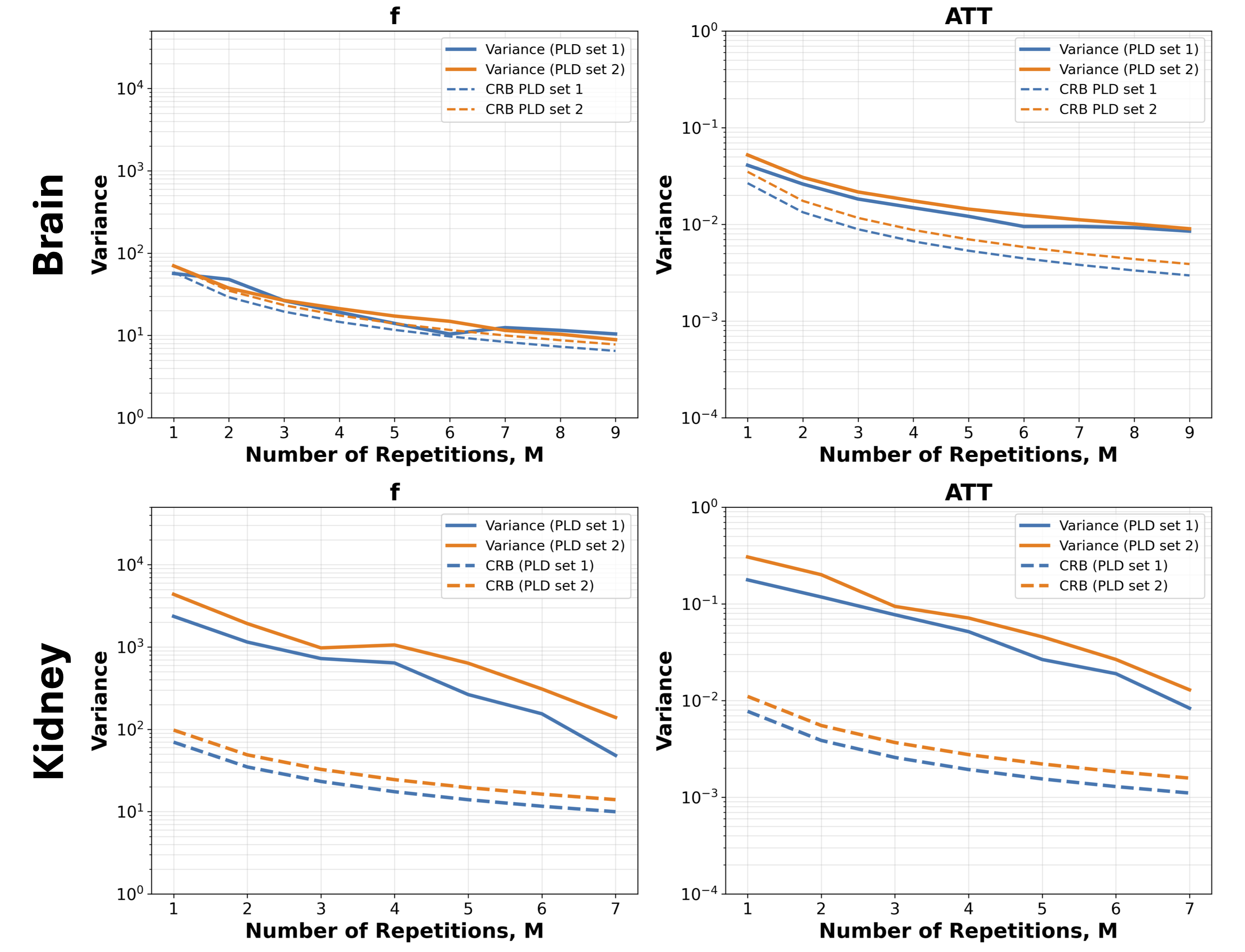}}
   \caption{Empirical MLE variance compared with the theoretical CRB (dotted lines). In the brain, the variance of $f$ remains tightly bounded by the CRB, and $\mathrm{ATT}$ shows a similar but slightly looser bound. In contrast, in the kidney the empirical variance exceeds the CRB by a larger margin, indicating that the CRB underestimates the estimation uncertainty.} 
   \label{fig:bounds}
 \end{figure}

 \begin{figure}[htbp]
   \centering
   \centerline{\includegraphics[width=\linewidth]{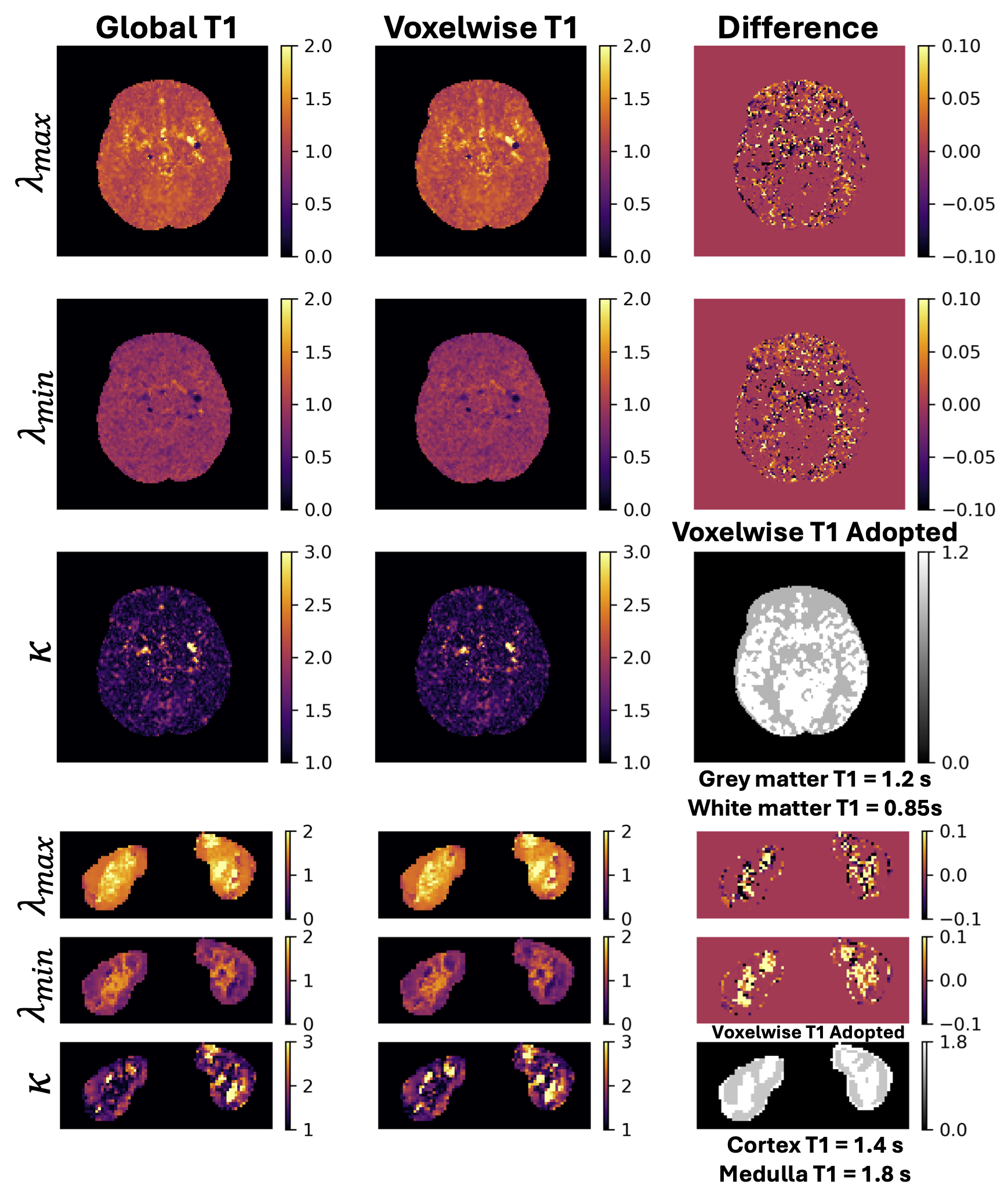}}
   \caption{Eigenvalues and condition number obtained with a global $T_1$ (left), a voxelwise $T_1$ (middle), and their difference (right). Voxelwise $T_1$ lowers both eigenvalues, yet the condition number remains large, suggesting that fixed $T_1$ and model mismatch both contribute to misspecification.} 
   \label{fig:t1}
 \end{figure}
 
\section{Results}

    Fig.~\ref{fig:converge} summarizes the simulated and \emph{in vivo} asymptotic convergence. In simulation, as the number of repetitions increases, both bias and variance decay toward 
    0, and the eigenvalues approach 1 from above and below. For the \emph{in vivo} data, the gap between $\lambda_{\max}$ and $\lambda_{\min}$ plateaus at values larger than simulation predicts.
    Fig.~\ref{fig:consist} shows the subset estimate consistency results. Relative errors in the kidney are larger and exhibit greater spatial heterogeneity.
    Fig.~\ref{fig:bounds} compares empirical variances with the theoretical CRB on a logarithmic scale. In the kidney, the empirical variance deviates from the CRB and is much less tightly bounded.
    Fig.~\ref{fig:t1} examines fixed parameter misspecification by replacing the global tissue $T_{1}$ with voxelwise values. Although both eigenvalues decrease with voxelwise $T_1$, the large condition number indicates that misspecification arises from both the fixed $T_1$ assumption and the underlying model.

\section{Discussion and Conclusion}
\label{sec:discussion}
This study presents a general framework for evaluating model misspecification in quantitative MRI. We assess misspecification through two criteria: the asymptotic convergence of variance bounds, and the consistency of parameter estimates. We also isolate model errors from fixed parameter misspecification. Using the ASL Buxton model as an case, our results suggest that the Buxton model appears to be specified in the brain, but may be moderately misspecified in the kidney. Future work will expand validation to larger cohorts and organs, and enable model selection across candidate qMRI models.
 
Many investigators have questioned the accuracy of the Buxton model assumptions in accurately capturing quantitative perfusion, particularly in the presence of disease. This work provides a data driven approach of assessing the validity of the model's assumptions, even in the absence of ground truth measurements \cite{fan2017long, bush2018pseudo}.

One plausible explanation for the Buxton model's misspecification in the kidney is because it assumes that labeled blood entering a voxel does not flow out and undergoes an exponential $T_1$ decay. In the kidney, however, labeled blood can exit the imaging voxel via efferent arterioles. This unmodeled outflow accelerates signal decay at long PLDs and leads to systematic underestimation of perfusion, as reported in \cite{alhummiany2022bias}.  
Partial-volume effects may also contribute to misspecification, as each ASL voxel often contains a mixture of compartments with distinct perfusion properties \cite{chappell2021partial}. 
We also find that using a fixed global tissue $T_1$ introduces additional error. Prior work \cite{bladt2020costs} has shown that jointly estimating $T_1$ in ASL compromises perfusion accuracy. Therefore, an additional $T_1$ mapping acquisition is recommended if feasible.


Our approach provides a global metric for model misspecification based on the condition number of $\hat{\mathbf{P}}$. Other discrepancy measures such as the matrix trace and KL divergence could also be used. Our work is also closely related to the information matrix and Hausman test proposed in \cite{white1982maximum}. While we focused on global measures, these tests could be applied per-voxel, though spatial correlations and multiple-comparison testing would need to be taken into account.


\section{Compliance with ethical standards}
\label{sec:ethics}
All data in this study were acquired under institutional review board (IRB) approval and informed consent.

\section{Acknowledgements}
This work is supported by NSF CCF-2239687 (CAREER) and the terminated NIH grant R01-EB033916.

\bibliographystyle{IEEEbib}
\bibliography{strings,refs}

@article{gudbjartsson1995rician,
  title={The Rician distribution of noisy MRI data},
  author={Gudbjartsson, H{\'a}kon and Patz, Samuel},
  journal={Magnetic resonance in medicine},
  volume={34},
  number={6},
  pages={910--914},
  year={1995},
  publisher={Wiley Online Library}
}

@article{alsop2015recommended,
  title={Recommended implementation of arterial spin-labeled perfusion MRI for clinical applications: a consensus of the ISMRM perfusion study group and the European consortium for ASL in dementia},
  author={Alsop, David C and Detre, John A and Golay, Xavier and G{\"u}nther, Matthias and Hendrikse, Jeroen and Hernandez-Garcia, Luis and Lu, Hanzhang and MacIntosh, Bradley J and Parkes, Laura M and Smits, Marion and others},
  journal={Magnetic resonance in medicine},
  volume={73},
  number={1},
  pages={102--116},
  year={2015},
  publisher={Wiley Online Library}
}

@article{odudu2018arterial,
  title={Arterial spin labelling MRI to measure renal perfusion: a systematic review and statement paper},
  author={Odudu, Aghogho and Nery, Fabio and Harteveld, Anita A and Evans, Roger G and Pendse, Douglas and Buchanan, Charlotte E and Francis, Susan T and Fern{\'a}ndez-Seara, Mar{\'\i}a A},
  journal={Nephrology Dialysis Transplantation},
  volume={33},
  number={suppl\_2},
  pages={ii15--ii21},
  year={2018},
  publisher={Oxford University Press}
}

@article{fortunati2017performance,
  title={Performance bounds for parameter estimation under misspecified models: Fundamental findings and applications},
  author={Fortunati, Stefano and Gini, Fulvio and Greco, Maria S and Richmond, Christ D},
  journal={IEEE Signal Processing Magazine},
  volume={34},
  number={6},
  pages={142--157},
  year={2017},
  publisher={IEEE}
}

@techreport{vuong1986cramer,
  title={Cram{\'e}r-Rao bounds for misspecified models},
  author={Vuong, Quang H.},
  year={1986},
  institution={Social Science Working Paper 652, California Institute of Technology},
  pages={1--38}
}

@article{white1982maximum,
  title={Maximum likelihood estimation of misspecified models},
  author={White, Halbert},
  journal={Econometrica: Journal of the Econometric Society},
  pages={1--25},
  year={1982},
  publisher={JSTOR}
}

@article{buxton1998general,
  title={A general kinetic model for quantitative perfusion imaging with arterial spin labeling},
  author={Buxton, Richard B. and Frank, Lawrence R. and Wong, Eric C. and Siewert, Bettina and Warach, Steven and Edelman, Robert R.},
  journal={Magnetic Resonance in Medicine},
  volume={40},
  number={3},
  pages={383--396},
  year={1998},
  publisher={Wiley Online Library}
}

@article{alhummiany2022bias,
  title={Bias and precision in magnetic resonance imaging-based estimates of renal blood flow: Assessment by triangulation},
  author={Alhummiany, Bandar A. and Shelley, Daniel and Saysell, Mark and Olaru, Maria A. and K{\"u}hn, Benjamin and Buckley, Daniel L. and Bailey, Jon and Wroe, Keith and Coupland, Carol and Mansfield, Michael W. and Sourbron, Steven P.},
  journal={Journal of Magnetic Resonance Imaging},
  volume={55},
  number={4},
  pages={1241--1250},
  year={2022},
  publisher={Wiley Online Library}
}

@article{parkes2002improved,
  title={Improved accuracy of human cerebral blood perfusion measurements using arterial spin labeling: accounting for capillary water permeability},
  author={Parkes, Laura M. and Tofts, Paul S.},
  journal={Magnetic Resonance in Medicine},
  volume={48},
  number={1},
  pages={27--41},
  year={2002},
  publisher={Wiley Online Library}
}

@article{bladt2020costs,
  title={The costs and benefits of estimating T1 of tissue alongside cerebral blood flow and arterial transit time in pseudo-continuous arterial spin labeling},
  author={Bladt, Peter and den Dekker, Arnold J. and Clement, Philippe and Achten, Eric and Sijbers, Jan},
  journal={NMR in Biomedicine},
  volume={33},
  number={12},
  pages={e4182},
  year={2020},
  publisher={Wiley Online Library}
}

@article{loh2008residual,
  title={Residual analysis for detecting mis-modeling in fMRI},
  author={Loh, Ji Meng and Lindquist, Martin A and Wager, Tor D},
  journal={Statistica Sinica},
  pages={1421--1448},
  year={2008},
  publisher={JSTOR}
}

@article{abed2021misspecified,
  title={Misspecified Cramer--Rao bounds for blind channel estimation under channel order misspecification},
  author={Abed-Meraim, Karim and Trung, Nguyen Linh and others},
  journal={IEEE Transactions on Signal Processing},
  volume={69},
  pages={5372--5385},
  year={2021},
  publisher={IEEE}
}

@article{cerreia2025making,
  title={Making decisions under model misspecification},
  author={Cerreia-Vioglio, Simone and Hansen, Lars Peter and Maccheroni, Fabio and Marinacci, Massimo},
  journal={Review of Economic Studies},
  pages={rdaf046},
  year={2025},
  publisher={Oxford University Press UK}
}

@article{fan2017long,
  title={Long-delay arterial spin labeling provides more accurate cerebral blood flow measurements in moyamoya patients: a simultaneous positron emission tomography/MRI study},
  author={Fan, Audrey P and Guo, Jia and Khalighi, Mohammad M and Gulaka, Praveen K and Shen, Bin and Park, Jun Hyung and Gandhi, Harsh and Holley, Dawn and Rutledge, Omar and Singh, Prachi and others},
  journal={Stroke},
  volume={48},
  number={9},
  pages={2441--2449},
  year={2017},
  publisher={Lippincott Williams \& Wilkins Hagerstown, MD}
}

@article{bush2018pseudo,
  title={Pseudo continuous arterial spin labeling quantification in anemic subjects with hyperemic cerebral blood flow},
  author={Bush, Adam and Chai, Yaqiong and Choi, So Young and Vaclavu, Lena and Holland, Scott and Nederveen, Aart and Coates, Thomas and Wood, John},
  journal={Magnetic resonance imaging},
  volume={47},
  pages={137--146},
  year={2018},
  publisher={Elsevier}
}

@article{chappell2021partial,
  title={Partial volume correction in arterial spin labeling perfusion MRI: A method to disentangle anatomy from physiology or an analysis step too far?},
  author={Chappell, Michael A and McConnell, Flora A Kennedy and Golay, Xavier and G{\"u}nther, Matthias and Hernandez-Tamames, Juan A and van Osch, Matthias J and Asllani, Iris},
  journal={Neuroimage},
  volume={238},
  pages={118236},
  year={2021},
  publisher={Elsevier}
}

\end{document}